\documentclass[twocolumn,showpacs,preprintnumbers,amsmath,amssymb]{revtex4}
\usepackage{amssymb}
\usepackage{graphicx}% Include figure files
\usepackage{dcolumn}% Align table columns on decimal point
\usepackage{bm}% bold math
\begin{document}

\title{Spin evolution of spin-1 Bose-Einstein condensates}

\author{Ma Luo, Zhibing Li, Chengguang
Bao\footnote{Corresponding author: stsbcg@mail.sysu.edu.cn}}

\address{The State Key Laboratory of Optoelectronic Materials and
Technologies \\ School of Physics and Engineering \\ Sun Yat-Sen
University, Guangzhou, 510275, P.R. China}

\begin{abstract} An analytical formula is obtained to describe the evolution of the average
populations of spin components of spin-1 \ atomic gases. The formula
is
derived from the exact time-dependent solution of the Hamiltonian $H_{S}=c%
\mathbf{S}^{2}$ without using approximation. \ Therefore it goes
beyond the mean field theory and provides a general, accurate, and
complete description for the whole process of non-dissipative
evolution starting from various initial states. \ The numerical
results directly given by the formula coincide qualitatively well
with existing experimental data, and also with other theoretical
results from solving dynamic differential equations. For some
special cases of initial state, instead of undergoing strong
oscillation as found previously, the evolution is found to go on
very steadily in a very long duration.
\end{abstract}


\maketitle

The liberation of the freedoms of spin of atoms in optical traps \cite%
{ho98,stam98,sten98,goel03,grie05} opens a new field, namely spin
dynamics of condensates, which is promising for super-high precise
measurement,
quantum computation, and quantum information processing. \cite%
{sore01,duan02,molm94} Recently, the evolution of spinor condensates
has
been extensively studied experimentally and theoretically. \cite%
{chang2004,law98,youli2005,pu99,diener2006} Initially, the
condensate was prepared in a Fock-state or a coherent state confined
in an optical trap. Then, due to the spin-dependent interaction, the
system begin to evolve where a pair of atoms with spin components 1
and -1 can jump to 0 an 0, and vice versa, via scattering. Finally
the system will arrive in equilibrium, however the process is not
smooth. In 1998, the average population of each of the spin
components $\mu =$1, 0, and -1 was found to depend sensitively on
initial states and may oscillate strongly with time. \cite{law98}.
This finding was further confirmed by a number of research groups.
In 2006, in
the study of the probability of finding a given number of bosons in a given $%
\mu $ state, the "quantum carpet" spin-time structure was found. \cite%
{diener2006} These findings show the amazing peculiarity of the spin
dynamics. Related theoretic calculations are mostly based on the
mean field theory. Although, in a number of particular cases,
theoretical results compares qualitatively well with experimental
data, the underlying physics remains to be further clarified. This
paper is a study of the evolution of the average populations. We
shall go beyond the mean field theory but use strict quantum
mechanic many-body theory with a full consideration of symmetry.
Instead of solving dynamic differential equations under specified
initial condition, we succeed to derive a general analytical formula
to describe rigorously the whole process of evolution
(non-dissipative) and is valid for all possible initial status. This
is reported as follows.

It is first assumed that the initial state of $N$ spin-1 atoms is a
Fock-state with populations $N_{1},N_{0}$ and $N_{-1}$, the magnetization $%
M=N_{1}-N_{-1}$. When $N$ and $M$ are given, the Fock-state can be
simply denoted as $|N_{0}\rangle $. Let the part of the Hamiltonian
responsible for
spin evolution be $H_{S}=c\mathbf{S}^{2}$, where $c$ is a constant, $\mathbf{%
S}$ is the operator of total spin. Then, the time evolution reads
\begin{equation}
\Psi (t)=e^{-iH_{S}t/\hbar }|N_{0}\rangle =\sum_{S}e^{-iS(S+1)\tau
}|\vartheta _{S,M}^{N}\rangle \langle \vartheta
_{S,M}^{N}|N_{0}\rangle \label{e1}
\end{equation}%
where $\tau =ct/\hbar $, and $|\vartheta _{S,M}^{N}>$ is the
all-symmetric total spin-state with good quantum numbers $S$ and
$M$. By using the analytical forms of the fractional parentage
coefficients and Clebesh-Gordan coefficients \cite{bao05,li06},
particle 1 can be extracted from the total spin-state as
\begin{eqnarray}
|\vartheta _{S,M}^{N}>= &&\sum_{\mathbf{\mu }}\chi _{\mathbf{\mu }%
}(1)[A(N,S,M,\mathbf{\mu })|\vartheta _{S+1,\ M-\mathbf{\mu
}}^{N-1}\rangle
\notag \\
&+&B(N,S,M,\mathbf{\mu })|\vartheta _{S-1,\ M-\mathbf{\mu
}}^{N-1}\rangle ] \label{e2}
\end{eqnarray}%
where $\chi _{\mathbf{\mu }}(1)$ is the spin-state of particle 1.
The coefficients involved in (\ref{e1}) and (\ref{e2}) are given in
the appendix. \ Inserting (\ref{e2}) into (\ref{e1}), the
probability of particle 1 in $\mathbf{\mu }$\ can be obtained, it
reads
\begin{equation}
\mathbf{P}_{N_{o},\mathbf{\mu }}^{M}(\tau )=\mathbf{B}_{N_{o},\mu }^{M}+%
\mathbf{O}_{N_{o},\mu }^{M}(\tau )  \label{e3}
\end{equation}%
where
\begin{equation}
\mathbf{B}_{N_{o},\mu }^{M}=\sum_{S}P_{\mathbf{\mu }}^{S,M}\ \langle
N_{0}|\vartheta _{S,M}^{N}\rangle \langle \vartheta
_{S,M}^{N}|N_{0}\rangle \label{e4}
\end{equation}%
\begin{equation}
P_{\mathbf{\mu }}^{S,M}=(A(N,S,M,\mathbf{\mu }))^{2}+(B(N,S,M,\mathbf{\mu }%
))^{2}  \label{e5}
\end{equation}%
\begin{equation}
\mathbf{O}_{N_{o},\mu }^{M}(\tau )=\sum_{S}O_{N_{o},\mu }^{M,S}\cos
(4(S+3/2)\tau )  \label{e6}
\end{equation}%
\begin{eqnarray}
O_{N_{o},\mu }^{M,S} &=&2A(N,S,M,\mathbf{\mu })B(N,S+2,M,\mathbf{\mu
})
\notag \\
&&\times \langle N_{0}|\vartheta _{S,M}^{N}\rangle \langle \vartheta
_{S+2,M}^{N}|N_{0}\rangle   \label{e7}
\end{eqnarray}%
The summation covers $S=N,\ N-2,\cdot \cdot \cdot \cdot \cdot
M^{\ast }$, where $M^{\ast }=M$ (or $M+1$) if $N-M$ is even (or
odd). Since the particles are identical, each of them plays the same
role, therefore the
average population in $\mathbf{\mu }$ is just $N\mathbf{P}_{N_{o},\mathbf{%
\mu }}^{M}(\tau )\equiv \langle a_{\mu }^{+}a_{\mu }\rangle $ (this
identity has been exactly proved numerically). \ In what follows
$\mathbf{\mu }=0$ is assumed (the cases with $\mathbf{\mu }\neq 0$
can be thereby understood). The label $\mathbf{\mu }$ may be
neglected from now on if $\mu =0$.

Eq.(\ref{e3}) is an exact consequence of the Hamiltonian $H_{S}=c\mathbf{S}%
^{2}$, no approximation has been introduced, it gives an analytical
description of the whole evolution (non-dissipative). There are time
dependent and independent terms, it implies an oscillation
surrounding a
background. It is straight forward from (\ref{e6}) that $\mathbf{P}%
_{N_{o}}^{M}(\mathbf{\tau })=\mathbf{P}_{N_{o}}^{M}(-\mathbf{\tau })=\mathbf{%
P}_{N_{o}}^{M}(\mathbf{\tau }+\pi )$, therefore $\mathbf{P}_{N_{o}}^{M}(%
\frac{\pi }{2}+\mathbf{\tau })=\mathbf{P}_{N_{o}}^{M}(\frac{\pi }{2}-\mathbf{%
\tau }).$It implies that the oscillation is periodic with the period
$\pi $
and $\ \mathbf{P}_{N_{o}}^{M}(\mathbf{\tau })$ is symmetric with respect to $%
\mathbf{\tau }=\frac{\pi }{2}$. Furthermore, since $\cos
(4(S+3/2)(\frac{\pi
}{4}+\mathbf{\tau }))=-\cos (4(S+3/2)(\frac{\pi }{4}-\mathbf{\tau })),$ $%
\mathbf{O}_{N_{o}}^{M}(\mathbf{\tau })$ is antisymmetric with respect to $%
\frac{\pi }{4}$, we have $\mathbf{P}_{N_{o}}^{M}(\frac{\pi
}{4}+\mathbf{\tau
})=2\mathbf{B}_{N_{o}}^{M}-\mathbf{P}_{N_{o}}^{M}(\frac{\pi }{4}-\mathbf{%
\tau }).$ Therefore, once $\mathbf{P}_{N_{o}}^{M}(\mathbf{\tau })$
has been known in the domain 0 to $\pi /4$, it can be known
everywhere. \ In
particular, $\mathbf{P}_{N_{o}}^{M}(0)=N_{0}/N$, $\ \mathbf{P}_{N_{o}}^{M}(%
\frac{\pi }{4})=\mathbf{B}_{N_{o}}^{M}$, and $\mathbf{P}_{N_{o}}^{M}(\frac{%
\pi }{2})=2\mathbf{B}_{N_{o}}^{M}-N_{0}/N$.

In $\mathbf{(}4\mathbf{)}$ the factor $P_{0}^{S,M}$ has an exact
analytical form as \cite{li06}
\begin{equation}
P_{0}^{S,M}=\frac{(2+1/N)S(S+1)-1-M^{2}(2+3/N)}{(2S+3)(2S-1)}
\label{e8}
\end{equation}%
When $N$\ is large, \ $P_{0}^{S,M}\approx \frac{1}{2}(1-(M/S)^{2}).$
Therefore,
\begin{equation}
\mathbf{B}_{N_{o}}^{M}\approx
\frac{1}{2}[1-\sum_{S}(\frac{M}{S})^{2}\langle N_{0}|\vartheta
_{S,M}^{N}\rangle \langle \vartheta _{S,M}^{N}|N_{0}\rangle ]\leq
\frac{1}{2}  \label{e9}
\end{equation}

In particular, when $M\rightarrow 0$, $\mathbf{B}_{N_{o}}^{M}\approx \frac{1%
}{2}$. The value 1/2 was first obtained numerically by Law, et al \cite%
{law98}$,$ and was supported by the recent study by Chang, et al \cite%
{chang2004}. Now this value is obtained analytically, and is further
found not depending on $N_{0}.$ When $M\rightarrow N,$ $S$ must also
tend to $N$, therefore both $P_{0}^{S,M}$ and
$\mathbf{B}_{N_{o}}^{M}\rightarrow 0$ as it should be.

For the time-dependent term, $O_{N_{o}}^{M,S}$ in (\ref{e6}) depends on $%
\mathbf{N}_{0}$\ strongly. There are three representative cases.

\begin{figure}
\includegraphics{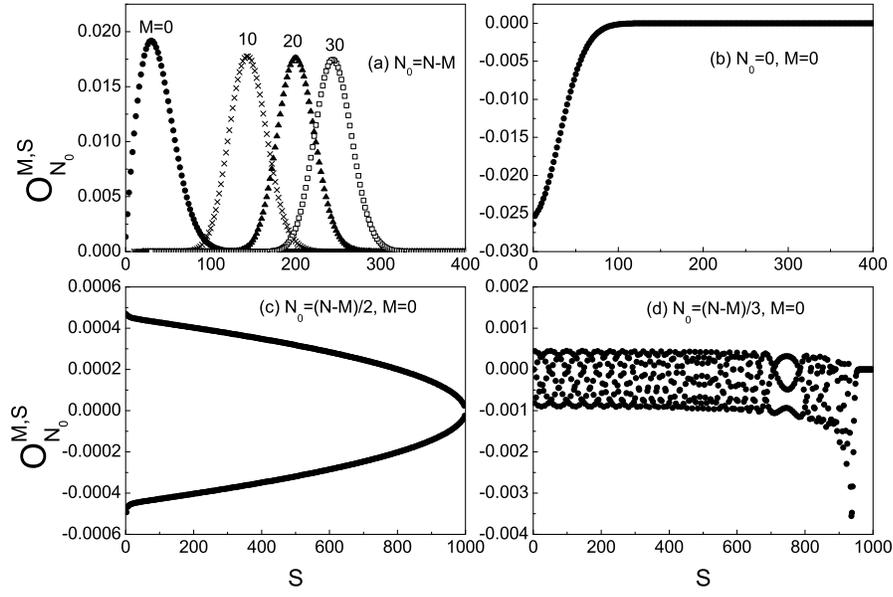}% Here is how to import EPS art
\caption{\label{o} $O^{M,S}_{N_{0}}$ versus $S$. $N=1000$ is given
(the same in the follows). }
\end{figure}

(i) When $N_{0}=N-M$ or $0$, $O_{N_{o}}^{M,S}$ is distributed in a
narrow
domain of $S$ (say, from $S_{a}$\ to $S_{b}$) as shown in Fig.\ref{o}a and %
\ref{o}b. \ In this case, when $O_{N_{o}}^{M,S}$ is roughly
considered as a constant in the narrow domain, from (\ref{e6}) we
have
\begin{equation}
\mathbf{O}_{N_{o}}^{M}(\mathbf{\tau })\approx \beta _{N_{o}}^{M}\overset{%
k_{\max }}{\underset{k=0}{\sum }}\cos (4(2k+S_{a}+3/2)\mathbf{\tau
})\equiv \beta _{N_{o}}^{M}G(\mathbf{\tau })  \label{e10}
\end{equation}
where $\beta _{N_{o}}^{M}$ is time-independent, $k=(S-S_{a})/2,$
$k_{\max }=(S_{b}-S_{a})/2$. $G(\mathbf{\tau })$ can be exactly
rewritten as
\begin{equation}
G(\mathbf{\tau })=\cos (4(S_{a}+3/2+k_{\max })\mathbf{\tau })\sin
(4(k_{\max }+1)\mathbf{\tau })/\sin (4\mathbf{\tau })  \label{e11}
\end{equation}
The denominator $\sin (4\mathbf{\tau })$ affects the behavior of
$G(\tau )$\ strongly. In the neighborhoods of $0$, the magnitude of
$G(\mathbf{\tau })$ would be remarkably larger because $\sin
(4\mathbf{\tau })$ is small, in particular, $G(0)=k_{\max }+1$. In
the neighborhoods of $\pi /4$, the magnitude of $G(\mathbf{\tau })$
would also be larger due to the denominator. However, since $G(\pi
/4)=0$, there would be a strong oscillation when $\tau \rightarrow
\pi /4$.

(ii) When $N_{0}\approx (N-M)/2$, $O_{N_{o}}^{M,S}$ is distributed
in a
broad domain of $S$ as shown in Fig.\ref{o}c where $O_{N_{o}}^{M,S}$ and $%
O_{N_{o}}^{M,S+2}$ have similar magnitudes but opposite signs. \ In
this case, the summation in (\ref{e6}) can be divided into two,
similarly we can define
\begin{eqnarray}
\overset{\sim }{G}(\tau ) &=&\overset{k_{\max }^{\prime }}{\underset{%
k^{\prime }=0}{\sum }}\cos (4(4k^{\prime }+S_{a}+3/2)\tau )  \notag \\
&&-\overset{k_{\max }^{\prime \prime }}{\underset{k^{\prime \prime
}=0}{\sum
}}\cos (4(4k^{\prime \prime }+S_{a}+7/2)\tau )  \notag \\
&=&\frac{2\sin (4\tau )}{\sin (8\tau )}\cdot \\
&&\sin (4(S_{a}+\frac{5}{2}+2k_{\max })\tau )\sin (8(k_{\max
}+1)\tau ) \notag  \label{e12}
\end{eqnarray}%
The feature of $\overset{\sim }{G}(\mathbf{\tau })$\ is greatly
different
from $G(\tau )$,\ in particular $\overset{\sim }{G}(0)=\overset{\sim }{G}%
(\pi /4)=0$, the denominator $\sin (8\mathbf{\tau })$ implies that $\overset{%
\sim }{G}(\mathbf{\tau })$ would be large in the neighborhood of $\mathbf{%
\tau }\approx \pi /8$. This leads to a very different feature of
evolution as shown later.

(iii) When $N_{0}$ is not close to the above cases, the variation of $%
O_{N_{o}}^{M,S}$ against $S$ has a band structure as shown in Fig.\ref%
{o}d, where neighboring $O_{N_{o}}^{M,S}$ and $O_{N_{o}}^{M,S+2}$
may have the same or opposite signs.

\begin{figure}
\includegraphics{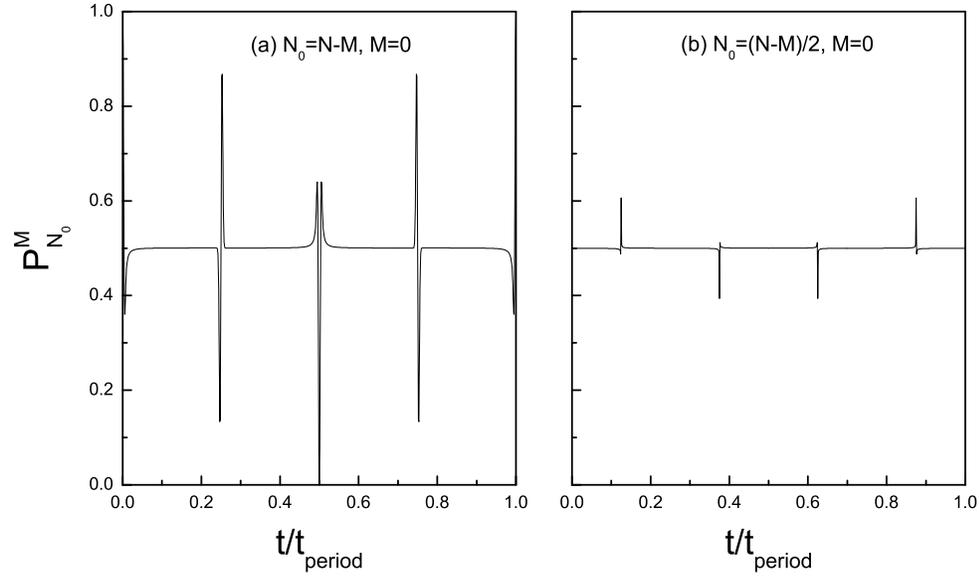}% Here is how to import EPS art
\caption{\label{t1} Evolution of the average population with $\mu
=0$. }
\end{figure}

Examples of $\mathbf{P}_{N_{o}}^{M}(\tau )$ calculated from
(\ref{e3}) are given in the follows. Fig.\ref{t1} shows the
evolution in the whole period 0 to $\pi $, where the strong
oscillation is concentrated in the neighborhoods of $k\pi /4$ (a) or
$k\pi /4+\pi /8$ (b), $k$\ is an integer, due to the
distinct features of $G(\mathbf{\tau })$ and $\overset{\sim }{G}(\mathbf{%
\tau })$. These figures show the symmetry in the period.
Experimentally, the duration of observation is much shorter than
$\pi .$ Evaluate under the Thomas-Fermi limit,\ when the trap is
described by an isotropic harmonic
potential with frequency $\omega /2\pi $, $\tau =\pi $ is associated with $%
t_{period}=\pi (N/\omega ^{2})^{3/5}X\sec $, where $X=1.52\times
10^{4}$ (3.86$\times 10^{3})$ for $^{87}$Rb ($^{23}$Na). In what
follows $\tau $\ is only given in a short duration.

\begin{figure}
\includegraphics{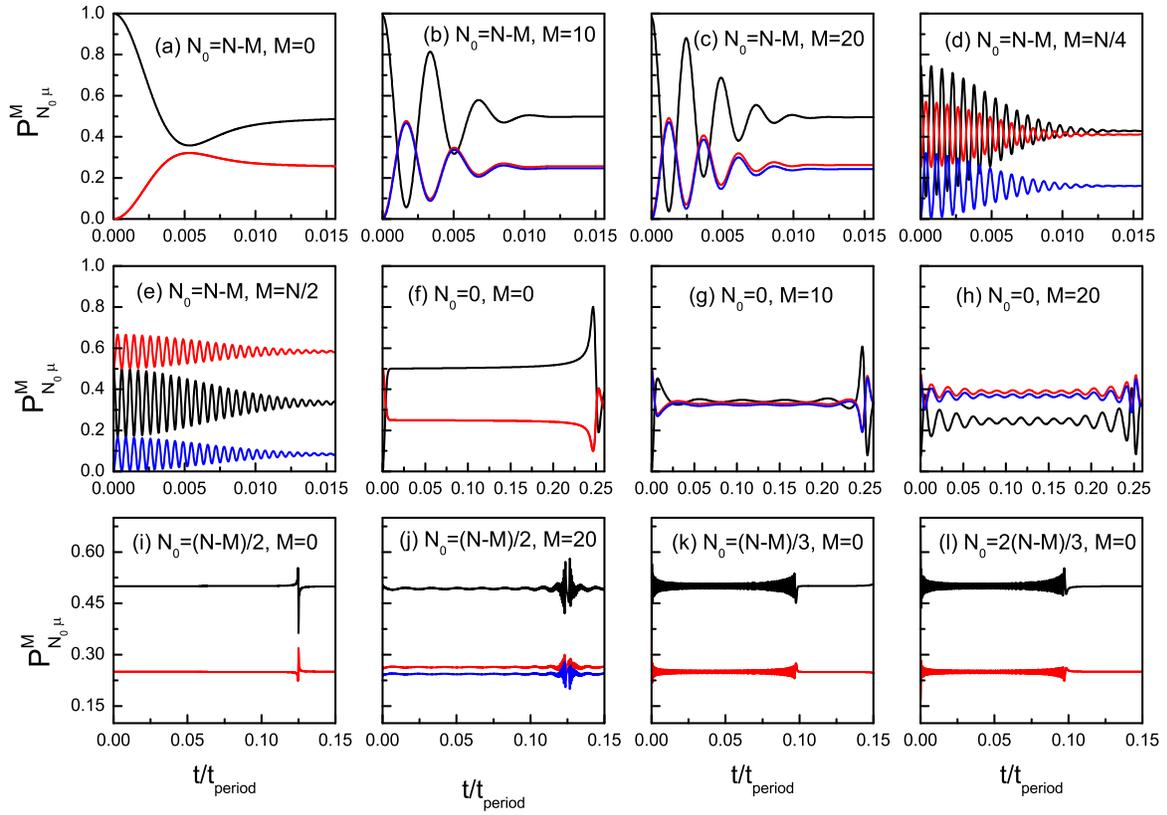}% Here is how to import EPS art
\caption{\label{t2} Evolution of the average populations with $\mu
=0$ (black), $1$ (red), and $-1$ (blue). For the case $M=0$, the red
and blue lines overlap. }
\end{figure}

The cases $N_{0}=N-M$ are shown in Fig.\ref{t2}a to \ref{t2}e. Fig.
\ref{t2}a is associated with the experiments by the MIT\ group
(upper panel of Fig.2 of \cite{sten98}); Fig. \ref{t2}b and c are
the cases that experiment error emerges which makes $M$ deviate from
$0$ slightly. Fig. \ref{t2}d and e are associated with the
experiments by GIT group (Fig.1 of \cite{youli2005} ), and
Hamburg group (Fig.5 of \cite{schm04}), respectively. Where, all\ $\mathbf{P}%
_{N_{o}}^{M}(\tau )$ (in solid lines)\ tend to
$\mathbf{B}_{N_{o}}^{M}=$ 1/2 or lower (if $M$ is larger) as
predicted above.

The cases $N_{0}=0$ are shown in Fig.\ref{t2}f to \ref{t2}h,
respectively.
Where \ref{t2}f is associated with the lower panel of Fig.2 of ref. \cite%
{sten98} [Stenger98].

The cases $N_{0}=(N-M)/2$ are shown in Fig.\ref{t2}i and \ref{t2}j. When $M$%
\ is small the evolution is very steady in a very long period 0 to
$\sim \pi /8$, then a strong oscillation occurs suddenly in the
neighborhood of $\pi /8 $ arising from the feature of $\overset{\sim
}{G}(\mathbf{\tau })$. Afterwards, the evolution becomes steady
again, and repeatedly.

When $N_{0}$ is not close to the above cases, two examples are given in Fig.\ref%
{t2}k and Fig.\ref{t2}l. The former one is the case discussed by Law, et al. (shown in Fig.3 of [\cite%
{law98}]). In this case, $O_{N_{o}}^{M,S}$ is nearly chaos (Fig.\ref{o}d), $%
\mathbf{P}_{N_{o}}^{M}(\tau )$ oscillates with $\tau $\ with a very
high frequency in the beginning, but suddenly disappears, and
suddenly recovers, and repeatedly.

In summary, this paper has essentially two findings

(1) Going beyond the mean field theory, without the necessity to
solve dynamical equations, a general analytical formula has been
derived based on
symmetry to describe the evolution of the average populations $\mathbf{P}%
_{N_{o}}^{M}(\tau )$ initiated from a pure Fock-state. This formula
is an exact consequence of the Hamiltonian $H_{S}=c\mathbf{S}^{2}$
with a full consideration of symmetry, no approximation is adopted.
Therefore the analysis based on this formula can help us \ to
understand better the peculiarity of spin evolution. For examples,
one can understand why\ the oscillation of
$\mathbf{P}_{N_{o}}^{M}(\tau )$ becomes very strong in somewhere (in
$\pi /4$ or $\pi /8$), why $\mathbf{P}_{N_{o}}^{M}(\tau )$ is
symmetric with respect to $\pi /2$, and so on. The results from the
formula coincides qualitatively with existing experimental data or
other theoretical results. It is expected that, when accurate
experimental data come out, a detailed quantitative comparison can
be made.

(2) A special initial state with $N_{0}=(N-M)/2$ and $M\approx 0$
was found where the evolution of $\mathbf{P}_{N_{o}}^{M}(\tau )$ is
steady in a very long duration from the begining until $\tau \approx
\pi /8$ \ This special stability is noticeable.

When the initial state is not a pure Fock-state but a superposition
of them, the generalization is straight forward.

\begin{acknowledgments} The support from the NSFC under the grants 10574163 and
90306016 are appreciated. \end{acknowledgments}

\section*{Appendix}
The coefficients in (\ref{e1}) and (\ref{e2}) are given as follows \cite%
{li06}
\begin{equation}
A(N,S,M,\mathbf{\mu })=a_{S}^{[N]}\;C_{1\mathbf{\mu },\;S+1,M-\mathbf{\mu }%
}^{S\;M}  \label{a1}
\end{equation}%
\begin{equation}
B(N,S,M,\mathbf{\mu })=b_{S}^{[N]}\;C_{1\mathbf{\mu },\;S-1,M-\mathbf{\mu }%
}^{S\;M}  \label{a2}
\end{equation}%
where
\begin{equation}
a_{S}^{[N]}=[(1+(-1)^{N-S})(N-S)(S+1)/(2N(2S+1))]^{1/2}  \label{a3}
\end{equation}%
\begin{equation}
b_{S}^{[N]}=[(1+(-1)^{N-S})\;S\;(N+S+1)/(2N(2S+1))]^{1/2}
\label{a4}
\end{equation}%
and $C_{1\mu ,\;S\pm 1,M-\mu }^{S\;M}$ are the Clebesh-Gorden
coefficients, their analytical forms are given in \cite{edmond}.

The set of coefficients $\langle \vartheta _{S,M}^{N}|N_{0}\rangle $
are obtained by diagonalizing the matrix of operator
$\hat{\mathbf{S}}^{2}$
\begin{equation}
\langle N_{b}^{\prime }|\hat{\mathbf{S}}^{2}|N_{b}\rangle
=A_{0}\delta _{N_{b}^{\prime },N_{b}}+A_{+}\delta _{N_{b}^{\prime
},N_{b}-2}+A_{-}\delta _{N_{b}^{\prime },N_{b}+2}  \label{a5}
\end{equation}%
where $A_{0}=M^{2}+N+N_{b}+2NN_{b}-2N_{b}^{2}$, $A_{+}=\sqrt{%
N_{b}(N_{b}-1)(N+M-N_{b}+2)(N-M-N_{b}+2)}$ and $A_{-}=\sqrt{%
(N_{b}+1)(N_{b}+2)(N+M-N_{b})(N-M-N_{b})}$.

\clearpage

\end{document}